\documentclass{article}
\usepackage{spconf,amsfonts,amsmath,amsthm,amssymb,graphicx}
\usepackage{adjustbox,lipsum,bm,soul}
\usepackage{cite}
\usepackage{color}
\usepackage{algorithm,algorithmic}
\usepackage{caption, subcaption}
\usepackage[hidelinks]{hyperref} 

\DeclareMathOperator*{\argmin}{arg\,min}
\DeclareMathOperator{\diag}{diag}
\DeclareMathOperator{\Diag}{Diag}

\newcommand{\Real}{{\mathbb{R}}}
\newcommand{\Complex}{{\mathbb{C}}}

\renewcommand{\vec}[1]{\ensuremath{\boldsymbol{#1}}}
\newcommand{\hvec}[1]{\ensuremath{\widehat{\boldsymbol{#1}}}}

\newcommand{\uvec}[1]{\ensuremath{\underline{\boldsymbol{#1}}}}
\newcommand{\tran}{{\mathsf{T}}}
\newcommand{\herm}{{\mathsf{H}}}

\newcommand{\defn}{\triangleq}
\newcommand{\figref}[1]{Fig.~\ref{fig:#1}}

\renewcommand{\eqref}[1]{(\ref{eq:#1})}

\usepackage{tikz}
\usetikzlibrary{fit, calc, shapes.geometric, arrows}

\title{Surface Coil Intensity Correction for MRI (Long Version)}
\name{Xuan Lei, Philip Schniter, Chong Chen, Muhammad Ahmad Sultan, and Rizwan Ahmad\thanks{Corresponding author: Rizwan Ahmad (ahmad.46@osu.edu).}}
\address{The Ohio State University}
\begin{document}
 \abovedisplayskip=3pt 
 \belowdisplayskip=3pt 
 \abovedisplayshortskip=3pt 
 \belowdisplayshortskip=3pt 
 \arraycolsep=3pt

\maketitle
\begin{abstract}
Modern MRI scanners utilize one or more arrays of small receive-only coils to collect k-space data. The sensitivity maps of the coils, when estimated using traditional methods, differ from the true sensitivity maps, which are generally unknown. Consequently, the reconstructed MR images exhibit undesired spatial variation in intensity. These intensity variations can be at least partially corrected using pre-scan data. In this work, we propose an intensity correction method that utilizes pre-scan data. For demonstration, we apply our method to a digital phantom, as well as to cardiac MRI data collected on a commercial scanner by Siemens Healthineers. The code is available at \url{https://github.com/OSU-MR/SCC}.


\end{abstract}

\section{Introduction}
\label{sec:intro}
MRI is a well-established noninvasive diagnostic tool. 
To meet the demands of shorter scan times, finer resolution, and higher signal-to-noise ratio, MRI hardware as well as data collection and processing methods have evolved rapidly in the last three decades. 
In particular, MRI scanners now use arrays of receive-only surface coils for signal detection. 
The k-space data are collected simultaneously across all receive coils, facilitating higher k-space undersampling and thus shorter scan times.
However, the sensitivity maps of the receive coils are spatially varying and generally not known in advance. 
The commonly used methods to find sensitivity maps from the k-space data do not yield true sensitivity maps but estimates in which the voxel-wise square-root of the sum-of-squares of the coil sensitivity maps equals one. 
Although this constraint allows estimating a unique set of sensitivity maps, the resulting estimated maps are spatially modulated relative to the true maps. 
When these imperfect sensitivity-map estimates are used in any SENSE-based reconstruction method \cite{pruessmann1999sense}, the reconstructed images also exhibit spatially varying intensity, leading to incorrect contrast and overly bright and dark regions. 
This can be specifically problematic for thoracic and abdominal imaging in larger patients because the central area of the body can appear excessively dark compared to the tissue closer to the received coils.
Post-processing methods based on filtering or histogram manipulation have been proposed to suppress intensity variation but have not been widely adopted \cite{murakami1996intensity, decarli1996local}. Although modern commercial scanners utilize data from pre-scans to partially correct the intensity variation, these corrections are not typically employed in research settings. This leads to mismatch between clinical and research imaging results. 
This work presents an open-source implementation for a Surface Coil intensity Correction method (SCC) for data collected on Siemens scanners.

\section{Methods}
\subsection{Problem formulation}
In MRI, the data are collected in the spatial frequency domain, called k-space, simultaneously from $K\geq 1$ receive coils. The data measured from the $k^{\text{th}}$ receive coil is modeled as
\begin{align}
    \vec{y}^{(k)} = 
    \underbrace{\vec{P}\vec{F}\vec{S}^{(k)}}_{\displaystyle \triangleq \vec{A}^{(k)}}
    \vec{x} + \vec{\eta}^{(k)},
    \label{eq:model-one-coil}
\end{align}
where $\vec{y}^{(k)} \in \Complex^{M}$ is measured k-space data, $\vec{x} \in \Complex^{N}$ is a vectorized 2D or 3D image, $\vec{S}^{(k)}\in \Complex^{N\times N}$ is a diagonal matrix representing the sensitivity map, $\vec{F} \in \Complex^{N \times N}$ is the 2D or 3D discrete Fourier transform, $\vec{P}$ is the sampling matrix, which contains $M$ rows from an $N\times N$ identity matrix, and $\vec{\eta}^{(k)} \in \Complex^{M}$ is additive circularly symmetric Gaussian noise with variance $\sigma^2$. By vertically stacking $\vec{y}^{(k)}$, $\vec{A}^{(k)}$, and $\vec{\eta}^{(k)}$ into $\vec{y} \in \Complex^{KM}$, $\vec{A} \in \Complex^{KM \times N}$, and $\vec{\eta} \in \Complex^{KM}$, respectively, the data measured from all $K$ coils can be expressed as
\begin{align}
    \vec{y} = \vec{A}\vec{x} + \vec{\eta}.
    \label{eq:model-all-coils}
\end{align}

To estimate $\vec{x}$ in \eqref{model-all-coils}, the regularized least-square (RL) solution is a common choice, which yields
\begin{align}
    \vec{x}_{\text{RL}} = \argmin_{\vec{x}} \frac{1}{\sigma^2}\|\vec{A}\vec{x} - \vec{y} \|_2^2 + \mathcal{R}(\vec{x}),
    \label{eq:rls}
\end{align}
where the regularization term $\mathcal{R}(\vec{x})$ encourages $\vec{x}_{\text{RL}}$ to adhere to a prior belief about the set of true $\vec{x}$.

To construct $\vec{A}$ in \eqref{rls}, one must know the true sensitivity maps $\{\vec{S}^{(k)}\}_{k=1}^K$. In practice, the true sensitivity maps are unknown, and so one typically uses an estimate $\{\hvec{S}^{(k)}\}_{k=1}^K$ produced by ESPIRiT \cite{uecker2014espirit}, by the method of Walsh et al. \cite{walsh2000adaptive}, or by simply dividing individual coil images by the square-root of the sum-of-squares (SSoS) \cite{pruessmann1999sense}, leading to
\begin{align}
    \hvec{x} = \argmin_{\vec{x}}\frac{1}{\sigma^2}\|\hvec{A}\vec{x} - \vec{y} \|_2^2 + \mathcal{R}(\vec{x}),
    \label{eq:espirit-solution}
\end{align}
where $\hvec{A}={\big[{\hvec{A}^{(1)\tran}},\dots, {\hvec{A}^{(K)\tran}} \big]}^\tran$ and $\hvec{A}^{(k)} = \vec{P}\vec{F}\hvec{S}^{(k)}$.

With ESPIRiT, SSoS, or similar methods, the estimates are constrained to $\hvec{S}^\herm \hvec{S}=\vec{I}$ for $\hvec{S} = \big[\hvec{S}^{(1)\tran},\dots, \hvec{S}^{(K)\tran}\big]^\tran$. 
Because the true sensitivity maps generally do not obey $\vec{S}^\herm\vec{S} = \vec{I}$, the estimated ones do not match the truth.

In SSoS, $\hvec{S}$ are estimated by elementwise dividing each coil image $\vec{S}^{(k)}\vec{x}$ with the square-root of the sum-of-squares image. Therefore, there exists a diagonal matrix $\vec{G}  \in \Real^{N\times N}$ such that $\vec{S}^{(k)} = \hvec{S}^{(k)} \vec{G}$ holds for all non-zero voxels in the noise-less case. In the typical case that $\vec{G} \neq \vec{I}$, the recovered $\hvec{x}$ in \eqref{espirit-solution} incurs unwanted spatial variation compared to $\vec{x}$. This problem can be addressed by estimating $\vec{G}$ from pre-scan data and integrating that estimate, $\hvec{G}$, into the forward model.
In particular, we propose to replace $\hvec{S}^{(k)}$ in \eqref{espirit-solution} with the corrected map $\hvec{S}^{(k)} \hvec{G}$ for each $k\in\{1,\dots,K\}$, giving
\begin{align}
    \hvec{x}_G = \argmin_{\vec{x}} \frac{1}{\sigma^2}\|\hvec{A}\hvec{G}\vec{x} - \vec{y} \|_2^2 + \mathcal{R}(\vec{x}).
    \label{eq:sens-corrected} 
\end{align}
Note that \eqref{sens-corrected} corrects the sensitivity maps prior to the reconstruction of $\vec{x}$. 
We also propose an alternative approach that first reconstructs $\vec{x}$ using uncorrected sensitivity maps and then corrects the reconstructed image with an estimate, $\hvec{H}$, of the true intensity correction map $\vec{H} \in \Real ^{N\times N}$, i.e.,
\begin{align}
    \hvec{x}_H = \hvec{H}\hvec{x},
    \label{eq:im-corrected}
\end{align}
where $\vec{H}$ and $\hvec{H}$ are both diagonal.
In the following section, we describe our proposed ``SCC'' method of estimating $\vec{G}$ and $\vec{H}$ from a short pre-scan that is invariably performed right before the primary data acquisition stage.

\newlength{\valOne}
\setlength{\valOne}{0.0cm} 
\newlength{\valTwo}
\setlength{\valTwo}{1cm} 
\newlength{\valThree}
\setlength{\valThree}{0.5cm} 
\newlength{\valFour}
\setlength{\valFour}{1.5cm} 
\newlength{\valFive}
\setlength{\valFive}{2.1cm} 
\newlength{\valSix}
\setlength{\valSix}{2.5cm} 

\tikzstyle{smallBox} = [rectangle, minimum width=\valFive, text width=\valFour, minimum height=\valOne, text centered, draw=black, font=\small, fill=red!30]
\tikzstyle{mediumBox} = [rectangle, minimum width=\valSix, text width=\valFive, minimum height=\valOne, text centered, draw=black, font=\small]
\tikzstyle{largeBox} = [rectangle, minimum width=\valSix, text width=\valFive, minimum height=\valOne, text centered, draw=black, font=\small]
\tikzstyle{arrow} = [thick,->,>=stealth]

\begin{figure*}[!h]
    \centering
    \begin{subfigure}{\textwidth}
    \begin{tikzpicture}[node distance=\valTwo]
    \node (box1) [mediumBox, fill=gray!20] {3D body coil images, $\uvec{x}_{\text{b}}$};
    \node (box2) [mediumBox, right of=box1, xshift=\valSix] {coil combining};
    \node (box3) [mediumBox, right of=box2, xshift=\valSix, yshift=-\valOne] {correction map using \eqref{g-opt}};
    \node (box4) [mediumBox, right of=box3, xshift=\valSix, yshift=\valOne] {interpolation, $\mathcal{I}(\hvec{g}_s;\vec{q})$};
    \node (box5) [mediumBox, right of=box4, xshift=\valSix] {reconstruction using \eqref{sens-corrected}};
    
    \node (box6) [mediumBox, fill=gray!20, below of=box1, yshift=-\valThree] {3D surface coil images, $\uvec{x}_{\text{s}}$};
    \node (box7) [mediumBox, below of=box2, yshift=-\valThree] {coil combining};
    \node (box8) [mediumBox, fill=gray!20, below of=box4, yshift=-\valThree] {MRI data, $\vec{y}$, slice location, $\vec{q}$};
    
    \draw [arrow] (box1) -- (box2);
    \draw [arrow] (box2.east) -- ($(box3.west)+(0,\valOne)$) node[midway, above, font=\small] {$\vec{x}_{\text{bc}}$};
    \draw [arrow] ($(box3.east)+(0,\valOne)$) -- (box4.west) node[midway, above, font=\small] {$\hvec{g}_{\text{s}}$};
    \draw [arrow] (box4) -- (box5) node[midway, above, font=\small] {$\hvec{G}$};
    \draw [arrow] (box5) -- ++(\valFive,0) node[midway, above, font=\small] {$\hvec{x}_{G}$};
    
    \draw [arrow] (box6) -- (box7);
    \draw [arrow] (box7.east) -- ($(box7.east)!0!(box3.west)$) -| node[above, near start, font=\small] {$\vec{x}_{\text{sc}}$} (box3.south); 
    \draw [arrow] (box8) -- (box4) node[midway, left, font=\small] {$\vec{q}$};
    \draw [arrow] (box8.east) -- ($(box8.east)!0!(box5.west)$) -| node[above, near start, font=\small] {$\vec{y}$} (box5.south);

    \node [draw, dashed, fit=(box1) (box6), inner sep=3pt] (dashedBox) {};
    \node [anchor=center, font=\small, xshift=0pt, yshift=0pt] at (dashedBox.center) {pre-scan};
    \node[anchor=north west, font=\large, xshift=0pt, yshift=15pt] at (current bounding box.north west) {(a)};
    \end{tikzpicture}
    \end{subfigure} 
    \\ \vspace{0.2cm}
    \begin{subfigure}{\textwidth}
    \begin{tikzpicture}[node distance=\valTwo]
    \node (box1) [mediumBox, fill=gray!20] {3D body coil images, $\uvec{x}_{\text{b}}$};
    \node (box2) [mediumBox, right of=box1, xshift=\valSix] {coil combining};
    \node (box3) [mediumBox, right of=box2, xshift=\valSix, yshift=-\valOne] {correction map using \eqref{h-opt}};
    \node (box4) [mediumBox, right of=box3, xshift=\valSix, yshift=\valOne] {interpolation, $\mathcal{I}(\hvec{h}_s;\vec{q})$};
    \node (box5) [mediumBox, right of=box4, xshift=\valSix] {correction using \eqref{im-corrected}};
    
    \node (box6) [mediumBox, fill=gray!20, below of=box1, yshift=-\valThree] {3D surface coil images, $\uvec{x}_{\text{s}}$};
    \node (box7) [mediumBox, below of=box2, yshift=-\valThree] {coil combining};
    \node (box8) [mediumBox, fill=gray!20, below of=box4, yshift=-\valThree] {MRI data, $\vec{y}$, slice location, $\vec{q}$};
    \node (box9) [mediumBox, below of=box5, yshift=-\valThree] {an image reconstruction};
    
    \draw [arrow] (box1) -- (box2);
    \draw [arrow] (box2.east) -- ($(box3.west)+(0,\valOne)$) node[midway, above, font=\small] {$\vec{x}_{\text{bc}}$};
    \draw [arrow] ($(box3.east)+(0,\valOne)$) -- (box4.west) node[midway, above, font=\small] {$\hvec{h}_{\text{s}}$};
    \draw [arrow] (box4) -- (box5) node[midway, above, font=\small] {$\hvec{H}$};
    \draw [arrow] (box5) -- ++(\valFive,0) node[midway, above, font=\small] {$\hvec{x}_{H}$};
    
    \draw [arrow] (box6) -- (box7);
    \draw [arrow] (box7.east) -- ($(box7.east)!0!(box3.west)$) -| node[above, near start, font=\small] {$\vec{x}_{\text{sc}}$} (box3.south); 
    \draw [arrow] (box8) -- (box4) node[midway, left, font=\small] {$\vec{q}$};
    \draw [arrow] (box8) -- (box9) node[midway, above, font=\small] {$\vec{y}$};
    \draw [arrow] (box9) -- (box5) node[midway, left, font=\small] {$\hvec{x}$};

    \node [draw, dashed, fit=(box1) (box6), inner sep=3pt] (dashedBox) {};
    \node [anchor=center, font=\small, xshift=0pt, yshift=0pt] at (dashedBox.center) {pre-scan};
    \node[anchor=north west, font=\large, xshift=0pt, yshift=15pt] at (current bounding box.north west) {(b)};
    \end{tikzpicture}
    \end{subfigure} 
    \caption{Two flavors of the proposed SCC method. In (a), the correction is applied to the sensitivity maps that are then used for image reconstruction. In (b), the correction is applied to images that have already been reconstructed. The content of the shaded boxes is available in the raw data file.}
    \label{fig:layout}
\end{figure*}
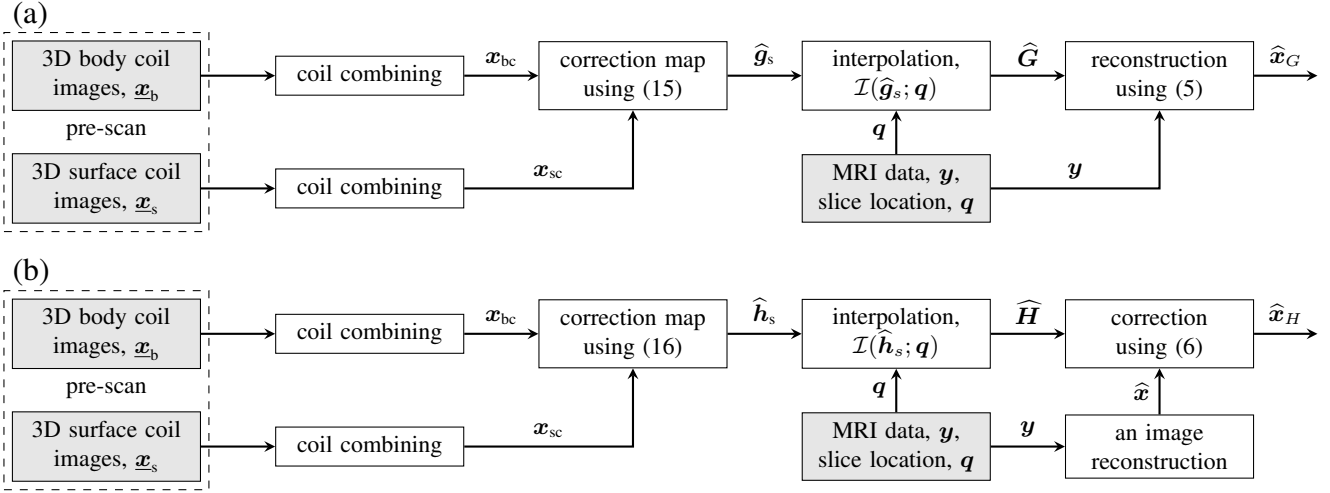

\subsection{Estimating correction maps from pre-scan}
Most clinical MRI scanners have two sets of RF coils: smaller surface receive coils that are placed close to the patient, and larger transmit/receive ``body coils'' that are hidden behind the bore liner. 
Since the advent of surface receive coils with high channel counts, body coils are primarily used for RF transmission. 
However, during the pre-scan, which is performed immediately before data acquisition, fully sampled low-resolution 3D images are simultaneously collected from surface and body coils. The main goal of this pre-scan is to facilitate the estimation of correction maps $\vec{G}$ or $\vec{H}$. 

Let $\vec{x}_{\text{3D}}\in \Complex^{L}$ denote the true, low-resolution 3D image with $L$ voxels, and let the diagonal matrix $\vec{S}_{\text{b}}^{(j)}\in \Complex^{L\times L}$ be the true sensitivity map of the $j^{\text{th}}$ body coil, with $j=1,2,\dots,J$. 
Then, the stack of true body-coil images $\uvec{x}_{\text{b}}\in \Complex^{JL}$ is given by $\uvec{x}_{\text{b}} = \vec{S}_{\text{b}} \vec{x}_{\text{3D}}$, where $\vec{S}_{\text{b}} = \big[\vec{S}_{\text{b}}^{(1)\tran}, \vec{S}_{\text{b}}^{(2)\tran},\dots, \vec{S}_{\text{b}}^{(J)\tran}\big]^\tran \in \Complex^{JL\times L}$. 

Let $\hvec{S}_{\text{b}}\in \Complex^{JL\times L}$ be the estimate of $\vec{S}_{\text{b}}$ constructed using SSoS. 
We assume that $\hvec{S}_{\text{b}}\approx\vec{S}_{\text{b}}$ because the body coils are large and have approximately uniform sensitivity maps. In this case, we can rewrite $\uvec{x}_{\text{b}} = \vec{S}_{\text{b}} \vec{x}_{\text{3D}}$ as
\begin{align}
    \uvec{x}_{\text{b}} &\approx \hvec{S}_{\text{b}} \vec{x}_{\text{3D}}
    \label{eq:body}.
\end{align}
By multiplying both sides of \eqref{body} with $\hvec{S}_{\text{b}}^\herm$, we get
\begin{align}
    \underbrace{\hvec{S}_{\text{b}}^\herm \uvec{x}_{\text{b}}}_{\displaystyle \defn \vec{x}_{\text{bc}}}
    &\approx
    \underbrace{\hvec{S}_{\text{b}}^\herm \hvec{S}_{\text{b}}}_{\displaystyle =\vec{I}}
    \vec{x}_{\text{3D}},
    \label{eq:body-approx}
\end{align}
which implies
\begin{align}
    \vec{x}_{\text{bc}} \approx \vec{x}_{\text{3D}}.
    \label{eq:body-3d}
\end{align}



Let the diagonal matrix $\vec{S}_{\text{s}}^{(k)}\in \Complex^{L\times L}$ be the true sensitivity map of the $k^{\text{th}}$ surface coil, with $k=1,2,\dots K$. Then the surface-coil image stack $\uvec{x}_{\text{s}}\in \Complex^{KL}$ is given by $\uvec{x}_{\text{s}} = \vec{S}_{\text{s}} \vec{x}_{\text{3D}}$, where $\vec{S}_{\text{s}} = \big[\vec{S}_{\text{s}}^{(1)\tran}, \vec{S}_{\text{s}}^{(2)\tran},\dots, \vec{S}_{\text{s}}^{(K)\tran}\big]^\tran \in \Complex^{KL \times L}$. 
If $\hvec{S}_{\text{s}}\in \Complex^{KL\times L}$ represents the coil maps estimated using SSoS, then there exists a diagonal $\vec{G}_{\text{s}} \in \Complex^{L\times L}$ such that $\vec{S}_{\text{s}}=\hvec{S}_{\text{s}}\vec{G}_{\text{s}}$ for all non-zero voxels in the noise-less case, in which case
\begin{align}
    \uvec{x}_{\text{s}} &= \hvec{S}_{\text{s}} \vec{G}_{\text{s}} \vec{x}_{\text{3D}} 
    \label{eq:surface} .
\end{align}
By multiplying both sides of \eqref{surface} with $\hvec{S}_{\text{s}}^\herm$, we get
\begin{align}
    \underbrace{\hvec{S}_{\text{s}}^\herm \uvec{x}_{\text{s}}}_{\displaystyle \defn \vec{x}_{\text{sc}}}
    &= 
    \underbrace{\hvec{S}_{\text{s}}^\herm \hvec{S}_{\text{s}}}_{\displaystyle =\vec{I}}
    \vec{G}_{\text{s}} \vec{x}_{\text{3D}},
    \label{eq:surface-approx}
\end{align}
which yields the coil-combined surface-coil image
\begin{align}
    \vec{x}_{\text{sc}} &= \vec{G}_{\text{s}}\vec{x}_{\text{3D}}. 
    \label{eq:surface-3d}
\end{align}
Combining \eqref{body-3d} and \eqref{surface-3d}, we get
\begin{align}
    \vec{x}_{\text{sc}} \approx \vec{G}_{\text{s}}\vec{x}_{\text{bc}}. 
    \label{eq:surface-body}
\end{align}
Since $\vec{G}_{\text{s}}$ is a diagonal matrix, 
\eqref{surface-body} can also be written as
\begin{align}
    \vec{x}_{\text{sc}} \approx \vec{X}_{\text{bc}}\vec{g}_{\text{s}}
    \label{eq:surface-body2}
\end{align}
with 
$\vec{X}_{\text{bc}} \defn \Diag(\vec{x}_{\text{bc}})\in\Real^{L\times L}$ and 
$\vec{g}_{\text{s}} \defn \diag(\vec{G}_{\text{s}}) \in \Real^{L}$.

We solve for $\vec{g}_{\text{s}}$ in \eqref{surface-body2} using regularized least-squares:
\begin{align}
    \hvec{g}_{\text{s}} = \argmin_{\vec{g}_{\text{s}}}\|\vec{X}_{\text{bc}}\vec{g}_{\text{s}}-\vec{x}_{\text{sc}}\|_2^2 + \phi(\vec{g}_{\text{s}}),
    \label{eq:g-opt}
\end{align}
where $\phi(\cdot)$ is a regularization term that promotes spatial smoothness. 
The 3D correction map $\hvec{g}_{\text{s}}$ can be interpolated along a 2D imaging plane to yield a 2D correction map $\hvec{g} = \mathcal{I}\big(\hvec{g}_{\text{s}};\vec{q}\big) \in \Real^N$, where $\vec{q}$ describes the orientation and position of the 2D imaging plane and $\mathcal{I}(\cdot; \vec{q})$ represents interpolation along that plane. Finally, $\hvec{g}$ can be embedded along the diagonal of $\hvec{G}$ and used in \eqref{sens-corrected} to obtain $\hvec{x}_G$. 

Alternatively, one can 
rewrite \eqref{surface-body} as $\vec{x}_{\text{bc}}\approx \vec{H}_{\text{s}}\vec{x}_{\text{bc}}$, with diagonal $\vec{H}_{\text{s}}\in\Real^{L\times L}$ and estimate 
$\vec{h}_{\text{s}}\defn\diag(\vec{H}_{\text{s}})$ via
\begin{align}
    \hvec{h}_{\text{s}} = \argmin_{\vec{h}_{\text{s}}}\|\vec{X}_{\text{sc}}\vec{h}_{\text{s}}-\vec{x}_{\text{bc}}\|_2^2 + \phi(\vec{h}_{\text{s}}),
    \label{eq:h-opt}
\end{align}
where $\vec{X}_{\text{sc}}\defn\Diag(\vec{x}_{\text{sc}})\in \Real^{L\times L}$. 
The 3D correction map $\hvec{h}_{\text{s}}$ can be interpolated along a 2D imaging plane to yield a 2D correction map $\hvec{h} = \mathcal{I}\big(\hvec{h}_{\text{s}};\vec{q}\big)\in \Real^N$. Then  
$\hvec{h}$ can be embedded along the diagonal of $\hvec{H}$ for use 
in \eqref{im-corrected} to yield $\hvec{x}_H$.

\figref{layout} describes the processing pipeline for producing corrected images $\hvec{x}_G$ and $\hvec{x}_H$ using SCC.

\begin{figure}[!t]
    \centering
    \includegraphics[width=0.8\linewidth]{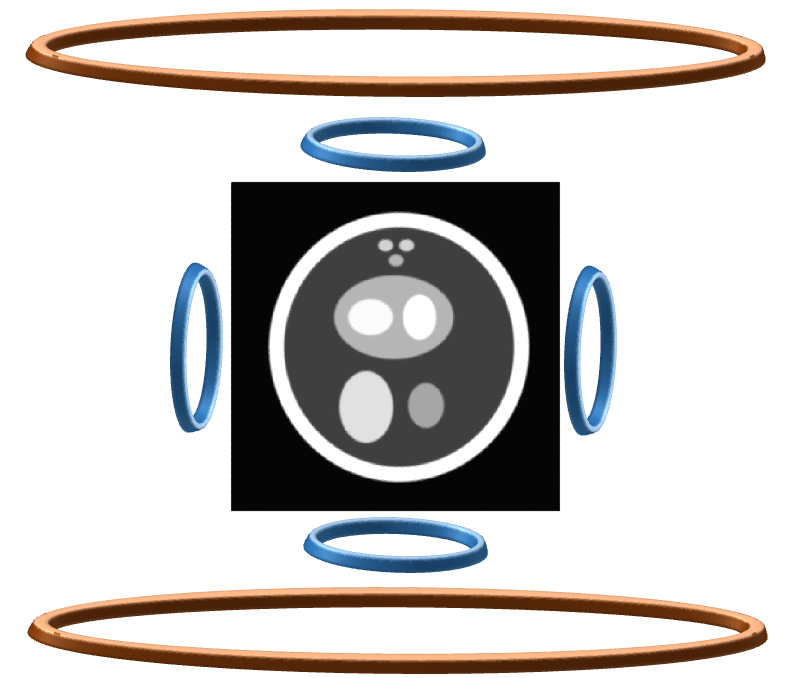}
    \put(-120,151){\makebox{\small body coil}}
    \put(-163,60){\rotatebox{90}{\small surface coil}}
    \caption{Layout of the digital phantom with four small surface coils and two large body coils.}
    \label{fig:phantom}
\end{figure}

\subsection{Implementation of SCC}
SCC was implemented with the regularization term $\phi(\cdot)\defn \lambda\|\left[\nabla_x^\tran(\cdot), \nabla_y^\tran(\cdot), \nabla_z^\tran(\cdot)\right]^\tran\|_2^2$, where $\nabla_x(\cdot)$, $\nabla_y(\cdot)$, and $\nabla_z(\cdot)$ compute first-order finite differences along the three directions of the input and $\lambda>0$ controls the regularization strength. 
Since both terms in \eqref{g-opt} and \eqref{h-opt} are quadratic, these optimization problems can be solved using closed-form expressions. Despite implementing $\vec{X}_{\text{bc}}$ and $\vec{X}_{\text{sc}}$ as sparse matrices, we observed that executing the closed-form solution was slow. 
In response, we solved the optimization problem in \eqref{g-opt} and \eqref{h-opt} using the conjugate gradient (CG) method. For 3D low-resolution images of size $64\times 64\times 64$, the CPU-only computation time for either of the optimization problems was less than 5 seconds on a standard workstation.

For real MRI data, $\uvec{x}_{\text{b}}$, $\uvec{x}_{\text{s}}$, $\vec{y}$, and $\vec{q}$ were extracted from the raw data file. After discarding oversampling in the readout direction, the field-of-view (FOV) and matrix size for $\uvec{x}_{\text{b}}$ and $\uvec{x}_{\text{s}}$ were $500\times 500 \times 500~\text{mm}^3$ and $64\times 32\times 32$, respectively, along the frequency encoding and two phase-encoding directions. 
The pre-scan data in the discrete Fourier domain were apodized with the Tukey window function to reduce Gibbs ringing and then zero-padded to deliver an isotropic resolution of $7.8\times 7.8 \times 7.8~\text{mm}^3$ and a matrix size of $64\times 64\times 64$. 
The multi-coil $\uvec{x}_\text{s}$ and $\uvec{x}_\text{b}$ images were coil-combined using SSoS, resulting in real-valued $\vec{x}_\text{sc}$ and $\vec{x}_\text{bc}$, respectively, of size $64\times 64 \times 64$. Both $\vec{x}_\text{sc}$ and $\vec{x}_\text{bc}$ were normalized by dividing with $\max\!\left[\vec{x}_\text{bc}\right]$ for \eqref{g-opt} and $\max\!\left[\vec{x}_\text{sc}\right]$ for \eqref{h-opt}. The value of $\lambda$ was set at $5\times 10^{-2}$. A Python implementation of SCC is available at \url{https://github.com/OSU-MR/SCC}.

\begin{figure}[!h]
    \begin{subfigure}{\linewidth}
		\includegraphics[width=\linewidth]{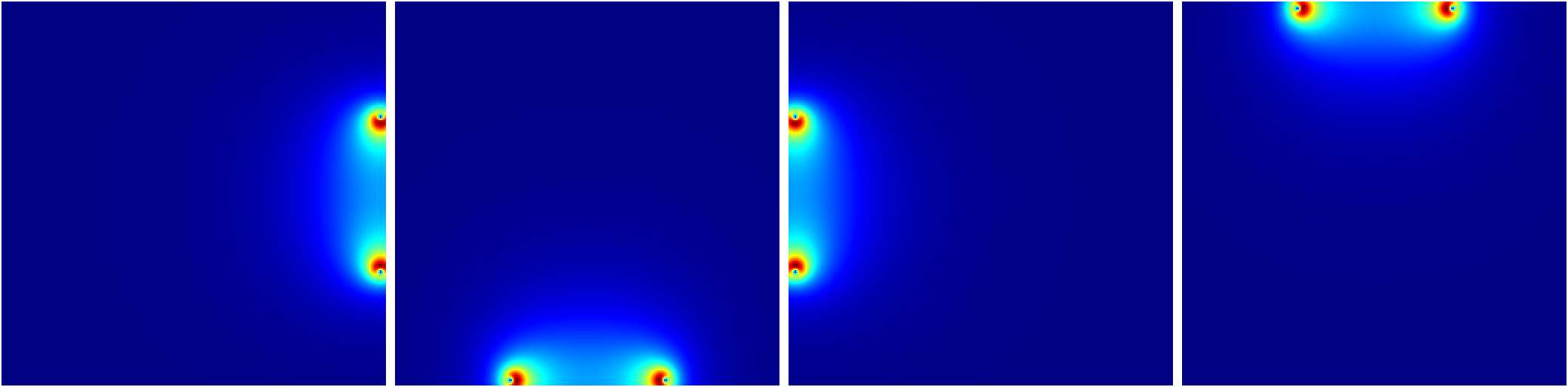}
        \put(-244,50){\colorbox{black}{\textcolor{white}{(a)}}}
    \end{subfigure}

    \vfill
    \begin{subfigure}{\linewidth}
		\includegraphics[width=\linewidth]{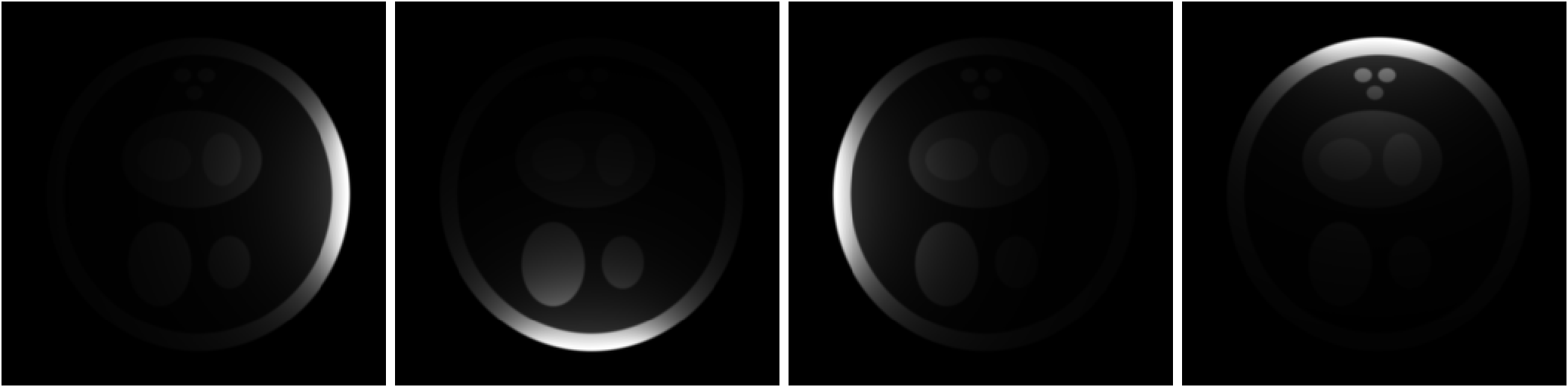}
        \put(-244,50){\colorbox{black}{\textcolor{white}{(b)}}}
    \end{subfigure}
    \vfill
    \begin{subfigure}{\linewidth}
		\includegraphics[width=\linewidth]{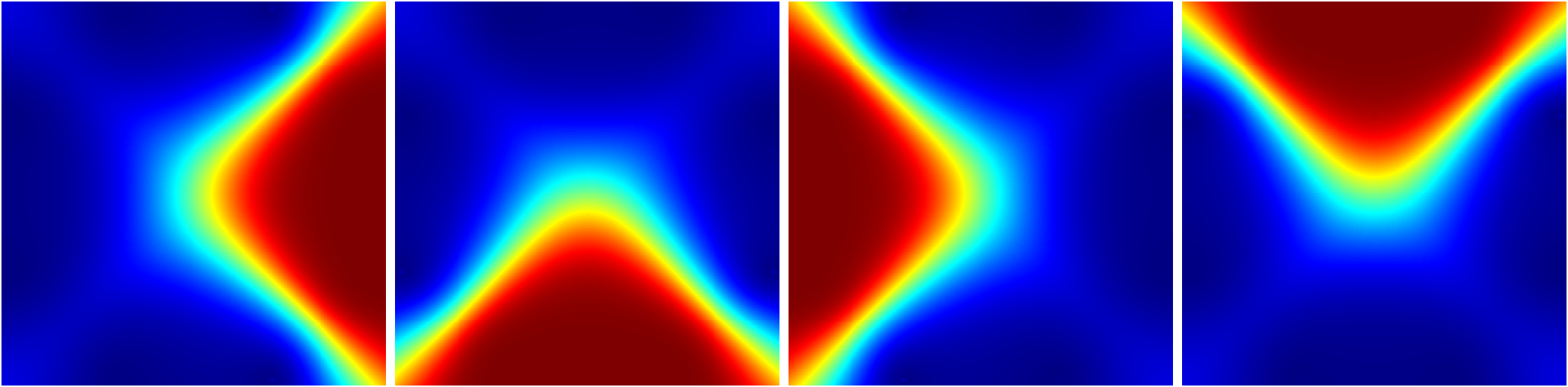}
        \put(-244,50){\colorbox{black}{\textcolor{white}{(c)}}}
    \end{subfigure}
    \vfill
    \begin{subfigure}{\linewidth}
		\includegraphics[width=\linewidth]{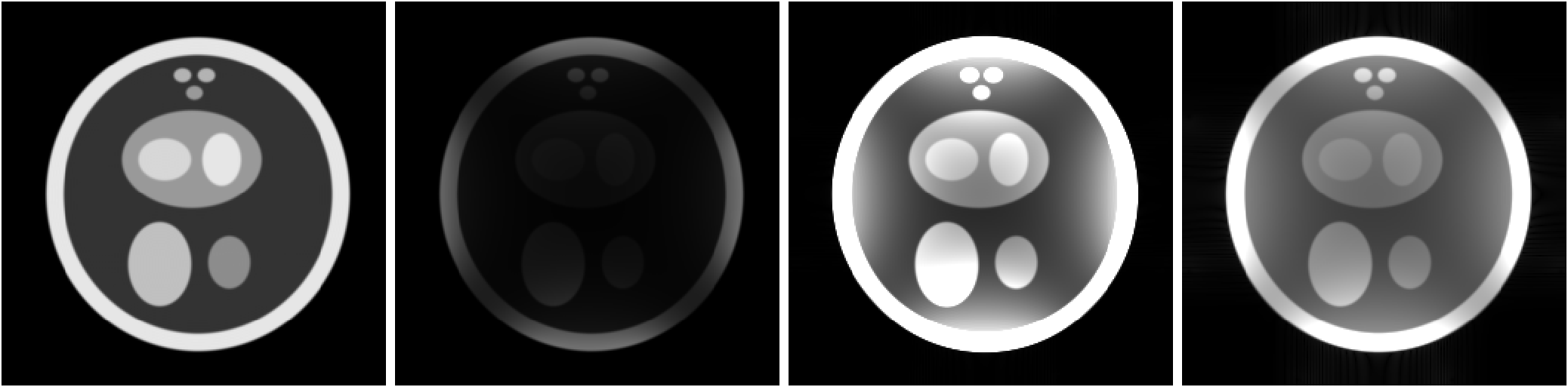}
        \put(-244,50){\colorbox{black}{\textcolor{white}{(d)}}}
    \end{subfigure}
    \vfill
    \begin{subfigure}{\linewidth}
		\includegraphics[width=\linewidth]{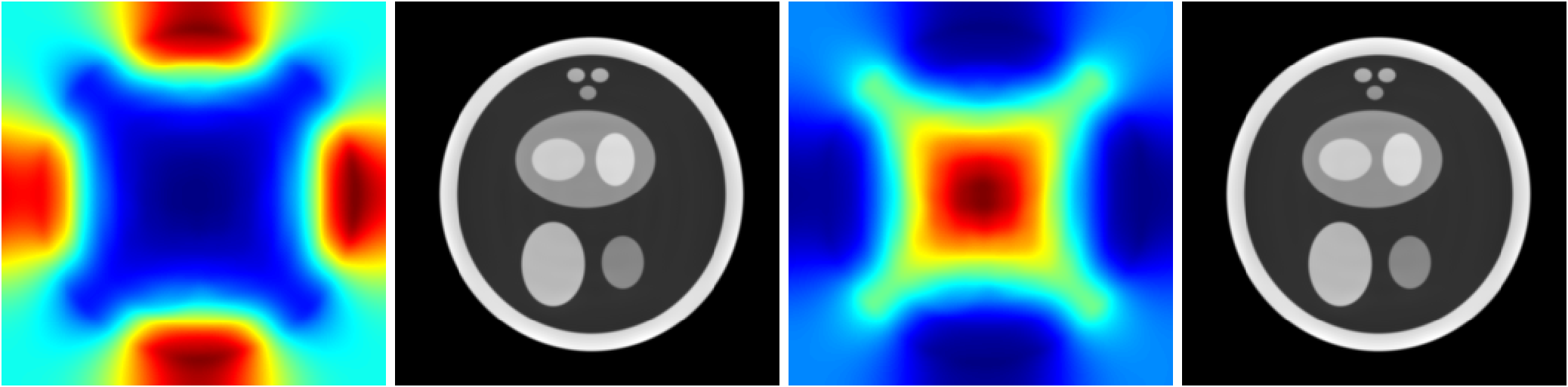}
        \put(-244,50){\colorbox{black}{\textcolor{white}{(e)}}}
    \end{subfigure}
     \caption{A simulation example showing the impact applying SCC. (a) True sensitivity maps simulated using the Biot-Savart law. (b) Element-wise multiplication of true sensitivity maps with a $256\times 256$ digital phantom. (c) Sensitivity maps estimated using SSoS. (d) From left to right: reference magnitude image $|\vec{x}|$, magnitude image $|\hvec{x}|$ recovered using SENSE with SSoS maps, $12\times |\hvec{x}|$, and $2\times |\hvec{x}|^{1/2}$. (e) From left to right: the correction map $\hvec{g}$, $|\hvec{x}_G|$, the correction map $\hvec{h}$, and $|\hvec{x}_H|$. All grayscale images are displayed on a scale of 0 to 1.}
     \label{fig:phantom-results}
\end{figure}

\begin{figure*}[ht]
    \centering
    \includegraphics[width=0.95\textwidth]{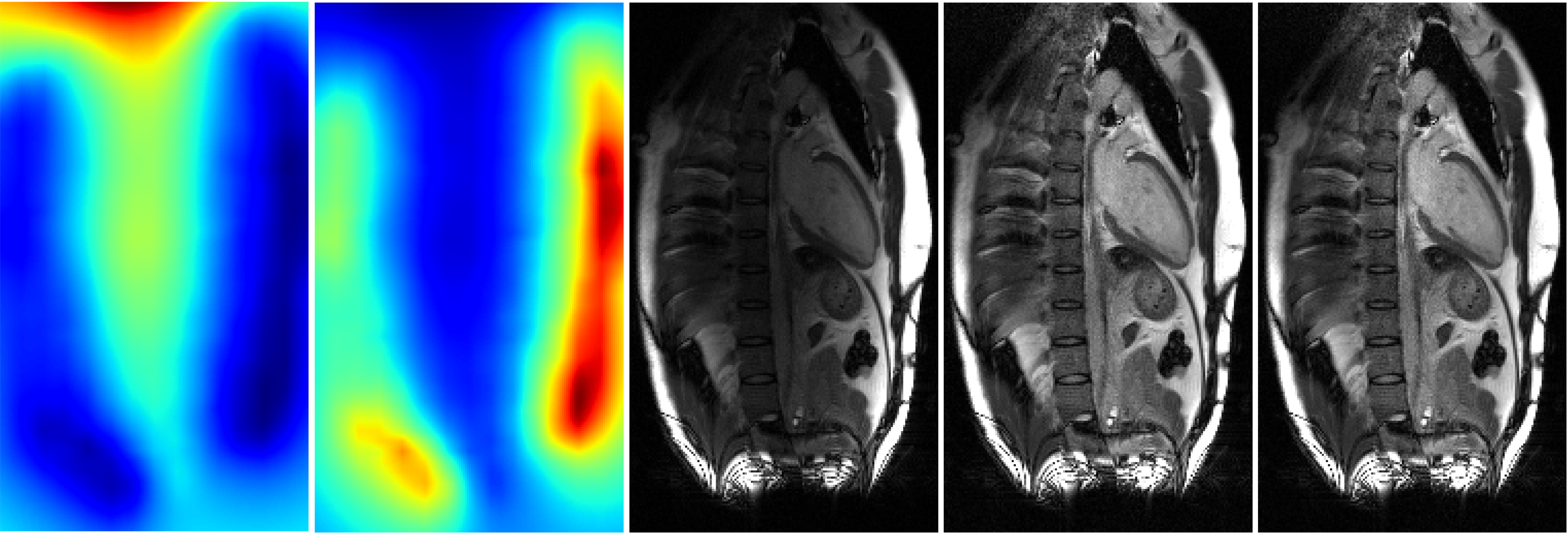}
    \put(-435,168){\textcolor{black}{\small $\hvec{g}$}}
    \put(-340,168){\textcolor{black}{\small $\hvec{h}$}}
    \put(-245,168){\textcolor{black}{\small $|\hvec{x}|$}}
    \put(-150,168){\textcolor{black}{\small $|\hvec{x}_G|$}}
    \put(-55,168){\textcolor{black}{\small $|\hvec{x}_H|$}}
    \caption{From left to right, two correction maps $\hvec{g}$ and $\hvec{h}$, the magnitude of the uncorrected image, and the magnitudes of the two corrected images. A representative two-chamber view of the heart is shown.}
    \label{fig:2ch}
\end{figure*}

\begin{figure*}[ht]
    \centering
    \includegraphics[width=0.95\textwidth]{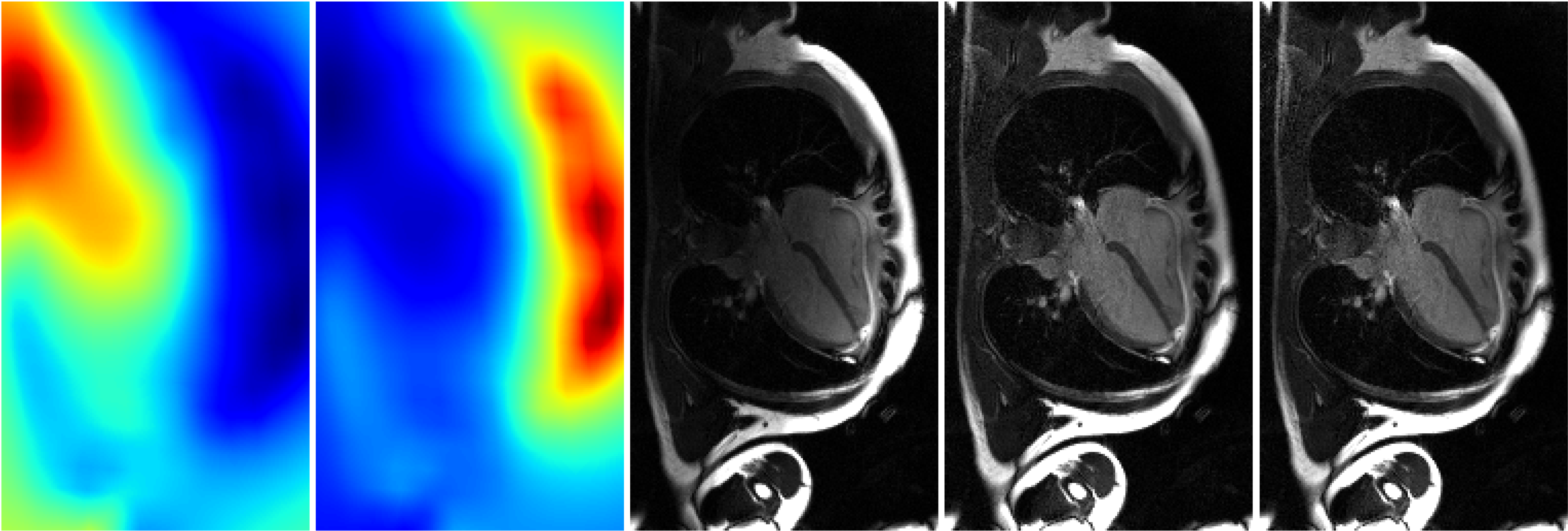}
    \put(-435,168){\textcolor{black}{\small $\hvec{g}$}}
    \put(-340,168){\textcolor{black}{\small $\hvec{h}$}}
    \put(-245,168){\textcolor{black}{\small $|\hvec{x}|$}}
    \put(-150,168){\textcolor{black}{\small $|\hvec{x}_G|$}}
    \put(-55,168){\textcolor{black}{\small $|\hvec{x}_H|$}}
    \caption{From left to right, two correction maps $\hvec{g}$ and $\hvec{h}$, the magnitude of the uncorrected image, and the magnitudes of the two corrected images. A representative four-chamber view of the heart is shown.}
    \label{fig:4ch}
\end{figure*}

\begin{figure*}[!ht]
    \centering
    \includegraphics[width=0.95\textwidth]{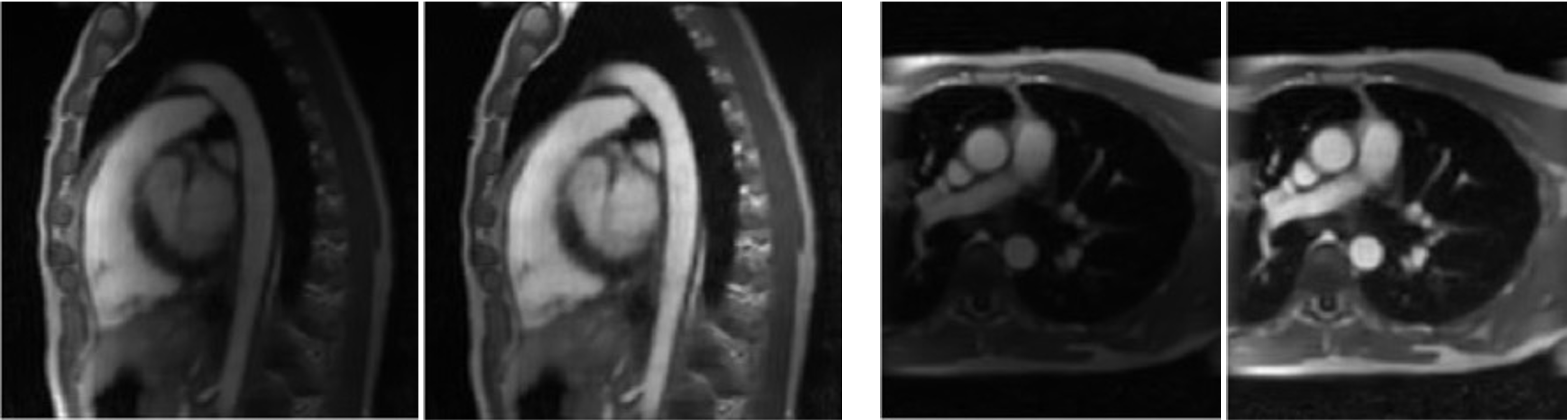}
    \put(-480,134){\textcolor{black}{\large (a)}}
    \put(-424,134){\textcolor{black}{\small $|\hvec{x}|$}}
    \put(-295,134){\textcolor{black}{\small $|\hvec{x}_H|$}}
    \put(-210,134){\textcolor{black}{\large (b)}}
    \put(-165,134){\textcolor{black}{\small $|\hvec{x}|$}}
    \put(-60,134){\textcolor{black}{\small $|\hvec{x}_H|$}}
    \caption{(a) A sagittal slice from a 4D flow image with ($|\hvec{x}_H|$) and without ($|\hvec{x}|$) correction, and (b) an axial slice from the same 4D flow image with ($|\hvec{x}_H|$) and without ($|\hvec{x}|$) correction.}
    \label{fig:4dflow}
\end{figure*}

\section{Experiment and Results}
\subsection{Simulation study}
In this study, we illustrate how SCC can improve the fidelity of the reconstructed image using a digital phantom. To this end, we simulated a $256 \times 256 $ image, $\vec{x}$, depicted in \figref{phantom}. 
The true sensitivity maps from four surface coils were generated using the Biot-Savart law. The coils were assumed to be circular wire loops, with a radius that is 20\% of the FOV, with the cylindrical axes of the coils pointing towards the center of the image. 
The true sensitivity maps were element-wise multiplied with $\vec{x}$ to generate individual coil images. Using SSoS as sensitivity maps, the image, $\hvec{x}$, was recovered using SENSE. 
As shown in the \figref{phantom-results}d, $\hvec{x}$ exhibits significant intensity variation, with the center of the image appearing dark. One could adjust the image intensity by histogram manipulation, but such measures do not fully address the problem. 
The true sensitivity maps (not shown) of two larger body coils, each with a radius equal to the FOV, were also generated using the Biot-Savart law. To simulate pre-scan, low-resolution data comprising $32 \times 32$ center of k-space were generated from both four small surface coils ($\uvec{x}_{\text{s}}$) and two large body coils ($\uvec{x}_{\text{b}}$). 
Following the procedure described in the previous section, two correction maps were constructed using SCC. However, the interpolation step was skipped because the digital phantom is already 2D. 
The last row in \figref{phantom-results} shows $\hvec{g}$, $\hvec{x}_G$, $\hvec{h}$, and $\hvec{x}_H$. Both $\hvec{x}_G$ and $\hvec{x}_H$ show marked improvement in the uniformity of image intensity. The normalized mean squared error (NMSE) defined as $\mathrm{NMSE}(\vec{x}, \hvec{x}) = 20\log_{10}\left(\|\vec{x} - \hvec{x}\|/\|\vec{x}\|\right)$ was $-5.67$ dB, $-27.64$ dB, and $-27.63$ dB for $\hvec{x}$, $\hvec{x}_G$, and $\hvec{x}_H$, respectively. 
The NMSE of $\hvec{x}_H$ or $\hvec{x}_G$ is primarily limited by the uniformity of $\uvec{x}_{\text{b}}$, which, in turn, depends on the body-coil size.

\subsection{Cardiac MRI study}
Single-shot free-breathing T1-weighted images were collected from a healthy volunteer on a 3T scanner (MAGNETOM Vida, Siemens Healthcare, Erlangen, Germany) running on an XA31A platform. 
The data were collected without contrast using phase-sensitive inversion recovery sequence \cite{kellman2002phase} and were prospectively undersampled at the acceleration rate of two. 
The raw data file, containing the pre-scan as well as the k-space measurements $\vec{y}$, was copied from the scanner for offline processing in Python. Representative results are shown in \figref{2ch} and \figref{4ch}. Both  $\hvec{x}_G$ and $\hvec{x}_H$ offer improvement over the uncorrected image $\hvec{x}$. In addition, ferumoxytol-enhanced 4D flow images were collected from a different subject on the same scanner using a five-minute free-running protocol \cite{pruitt2021fully}. Representative results are shown in \figref{4dflow}. In this case, the correction was applied to $\hvec{x}$ reconstructed using traditional compressed sensing \cite{lustig2007sparse} to yield $\hvec{x}_H$. However, it is also possible to estimate $\hvec{x}_G$ using \eqref{sens-corrected}. 

\subsection{Discussion}
The quality of SCC correction depends on the spatial homogeneity of $\vec{x}_{\text{bc}}$. 
Due to the much larger size of the body coils, one expects $\vec{x}_{\text{bc}}$ to be more uniform than $\uvec{x}_{\text{s}}$. In practice, however, $\vec{x}_{\text{bc}}$ may not offer the perfect uniformity, leading to residual intensity variation in $\hvec{x}_G$ and $\hvec{x}_H$. 
The results presented in \figref{phantom-results} and \figref{2ch} show little difference between $\hvec{x}_G$ and $\hvec{x}_H$. These reconstructions are based on SENSE without regularization. 
However, for deep learning-based methods or methods that use regularization, e.g., compressed sensing, we expect $\hvec{x}_G$ and $\hvec{x}_H$ to differ. 
This is because $\hvec{x}_G$ may have better spatial uniformity than $\hvec{x}$, which can potentially make $\hvec{x}_G$ more amenable to sparsity-based or generative priors. 

\section{Conclusions}
MRI reconstructions performed offline typically do not utilize the pre-scan data to normalize image intensity, leading to uneven image intensity, which can be especially problematic for abdominal and thoracic imaging from larger subjects. 
We propose an optimization-based method to correct surface coil intensity and provide an open-source Python implementation. Future efforts will include a demonstration of SCC in other MRI applications and with scanners from other vendors.

\section{Additional Information}
\textbf{Acknowledgment}: This work was funded by NIH grants R01EB029957 and R01HL151697.
\vspace{2mm}

\noindent
\textbf{Compliance with ethical standards}: This study was performed in line with the principles of the Declaration of Helsinki. Approval was granted by the Institutional Review Board (IRB) of The Ohio State University (2019H0076).



\bibliographystyle{IEEEbib}
\bibliography{master.bib}

\end{document}